\documentclass[aps,prl,showpacs,twocolumn,groupedaddress]{revtex4}
\usepackage{graphics}
\usepackage{graphicx}
\newcommand \beq{\begin{eqnarray}}
\newcommand \eeq{\end{eqnarray}}
\newcommand \bea{\begin{eqnarray}}
\newcommand \eea{\end{eqnarray}}
\newcommand \kvec{{\bf k}}
\newcommand \qvec{{\bf q}}
\newcommand \pvec{{\bf p}}
\newcommand \Rvec{{\bf R}}

\newcommand \rvec{{\bf r}}
\newcommand \ex{{\bf \hat{x}}}
\newcommand \ey{{\bf \hat{y}}}
\def\simge{\mathrel{%
      \rlap{\raise 0.511ex \hbox{$>$}}{\lower 0.511ex \hbox{$\sim$}}}}
\def\simle{\mathrel{
      \rlap{\raise 0.511ex \hbox{$<$}}{\lower 0.511ex \hbox{$\sim$}}}}

\begin{document}

   \title{Renormalization factor and effective mass of the two-dimensional
   electron gas}

\author{Markus Holzmann,$^{a,b}$, Bernard Bernu$^a$,  Valerio Olevano$^c$, Richard M. Martin$^{d,e}$, and David M. Ceperley$^e$
\\
$^a$LPTMC, UMR 7600 of CNRS, Universit\'e Pierre et Marie Curie, 4 Place Jussieu,
 Paris, France
\\
$^b$LPMMC, CNRS-UJF, UMR 7644 of CNRS,   BP 166,  Grenoble, France
\\
$^c$Institut N{\'e}el, CNRS-UJF,  Grenoble, France
\\
$^d$Department of Applied Physics, Stanford University, Stanford, CA 94305
\\
$^e$Department of Physics, University of Illinois, 1110 W. Green St.,
Urbana, IL 61801
}

\date{\today}

\begin{abstract}

We calculate the momentum distribution of the Fermi liquid
phase of the homogeneous, two-dimensional electron gas. We show
that, close to the Fermi surface, the momentum distribution of
a finite system with $N$ electrons approaches  its
thermodynamic limit slowly, with leading order corrections
scaling as $N^{-1/4}$. These corrections dominate the
extrapolation of the renormalization factor, $Z$,  and the
single particle effective mass, $m^*$, to the infinite system
size. We show how convergence can be improved analytically. In
the range $1 \le r_s \le 10$, we get a lower renormalization
factor  $Z$ and a higher effective mass, $m^*>m$, compared to
the perturbative RPA values.

\end{abstract}


\maketitle

Landau's Fermi liquid theory \cite{Landau} postulates a
one-to-one mapping of low energy excitations  of an interacting
quantum system with that of an ideal Fermi gas via the
distribution function of quasiparticles of momentum $k$. The
resulting energy functional has been successfully applied to
describe equilibrium and transport properties of quantum Fermi
liquids, the most prominent are the electron gas and liquid
$^3$He \cite{Pines,Gordon}. However, quantitative microscopic
calculations of its basic ingredients, the renormalization
factor, $Z$, and the effective mass, $m^*$, remain challenging.

In this paper, we calculate these parameters for the 2DEG
(two-dimensional electron gas) using Quantum Monte Carlo (QMC)
in the region $1 \le r_s \le 10$, where $r_s=(\pi n a_B^2)^{-1/2}$
is the Wigner-Seitz density parameter, $n$ the density, and
$a_B=\hbar^2/(me^2)$ the Bohr radius.
Kwon et al.\cite{Kwon1} made the first attempt to
determine the Fermi liquid parameters of the two-dimensional
electron gas using QMC,
with results that differ from
calculations based on
other methods \cite{Schuck,Giuliani,Vignale}.
In particular, Kwon et al. found an effective mass
smaller than the bare mass, \emph{e.g.} $m^*<m$ at $r_s=1$.
However, QMC calculations suffer from severe finite size effects
since typical  system sizes are limited to $N \sim 100$ electrons.
Here we show, that
there is an extremely slow convergence of
the effective mass and the renormalization factor
to their  thermodynamic limit values, with
leading order corrections scaling as $N^{-1/4}$. A correct
extrapolation to the infinite sized system leads to important
qualitative and quantitative differences compared to previous
calculations \cite{Kwon1, Vignale} which had assumed a ($1/N$)
extrapolation. We further use the knowledge of the analytical
properties of the ground state wavefunction \cite{FSE}, to
analytically estimate dominant and sub-dominant size effects which are important.

Microscopically, the existence and characteristics of the Fermi
surface of interacting fermions are directly related to the
renormalization factor $Z$ at the Fermi surface
\cite{Luttinger}. For a normal Fermi liquid, one expects a
sharp discontinuity in the momentum distribution, $n_\kvec$, at
the Fermi surface for each spin \beq n_{\kvec}=Z_\kvec
\theta(k_F-|\kvec |) +g_\kvec \label{nkdef} \eeq where $k_F$ is
the Fermi wavevector, $\theta(x)$ is a step-function, and
$g_\kvec$ is a continuous function of momentum $\kvec$.  One
expects a linear excitation spectrum $\varepsilon_\kvec=
\hbar^2 k_F  ( k-k_F)/m^*$ close to the Fermi surface at
$|\kvec|=k_F$, with $m^*$ the effective mass. The goal is to
determine the properties in the $N \rightarrow \infty$ limits
based upon calculations on cells of $N$ particles with discrete
values of $n_{\kvec}^N$ and $Z_\kvec^N$.  
A microscopic construction of the Landau energy functional
is based on considering energy eigenstates 
which are adiabatically connected to the excited states of  the non-interacting 
Fermi gas \cite{Vignale}.
The effective mass is then explicitly given in terms of  the energy difference,
$\varepsilon_\kvec=E_\kvec^{N+1}-(E_0^{N}+\mu)$,
between single particle excitations of energy $E_\kvec^{N+1}$ and 
momentum $\kvec$, and 
 the $N$-particle ground state of energy $E_0^N$ 
 where $\mu$ is the chemical potential.

Quantum Monte Carlo (QMC) methods provide the most accurate
calculations of the ground state energy of the electron gas
\cite{EG,Tanatar}. However, fermionic QMC calculations suffer
from two major drawbacks, the fixed node approximation and
finite size errors. For a normal Fermi liquid, the most precise
results are obtained using a generalized Slater Jastrow form
for the trial wavefunction \cite{he3} \beq \Psi_T \propto
D(\Rvec) e^{-U(\Rvec)} \label{psit} \eeq where $\Rvec$
indicates a dependance on all particle coordinates.
Antisymmetry is assured by a Slater determinant
$D(\Rvec)=\det_{ij} e^{i \kvec_j \cdot \qvec_i(\Rvec)} $ of
plane waves inside the Fermi sphere $|\kvec_j| \le k_F$ using
dressed quasiparticle coordinates $\qvec_i(\Rvec)$ to account
for many-body backflow effects, whereas the many-body Jastrow
potential $U(\Rvec)$ is symmetric with respect to particle
exchange and accounts for the singularities in the
interparticle potential at the coincidence points. Projector
Monte Carlo methods (DMC) can be used to improve the
wavefunction stochastically. Many  ground-state properties
have been successfully calculated using QMC, however, the
situation is less clear concerning excited state properties
\cite{Vignale}.

\begin{figure}
\includegraphics[width=0.45\textwidth]{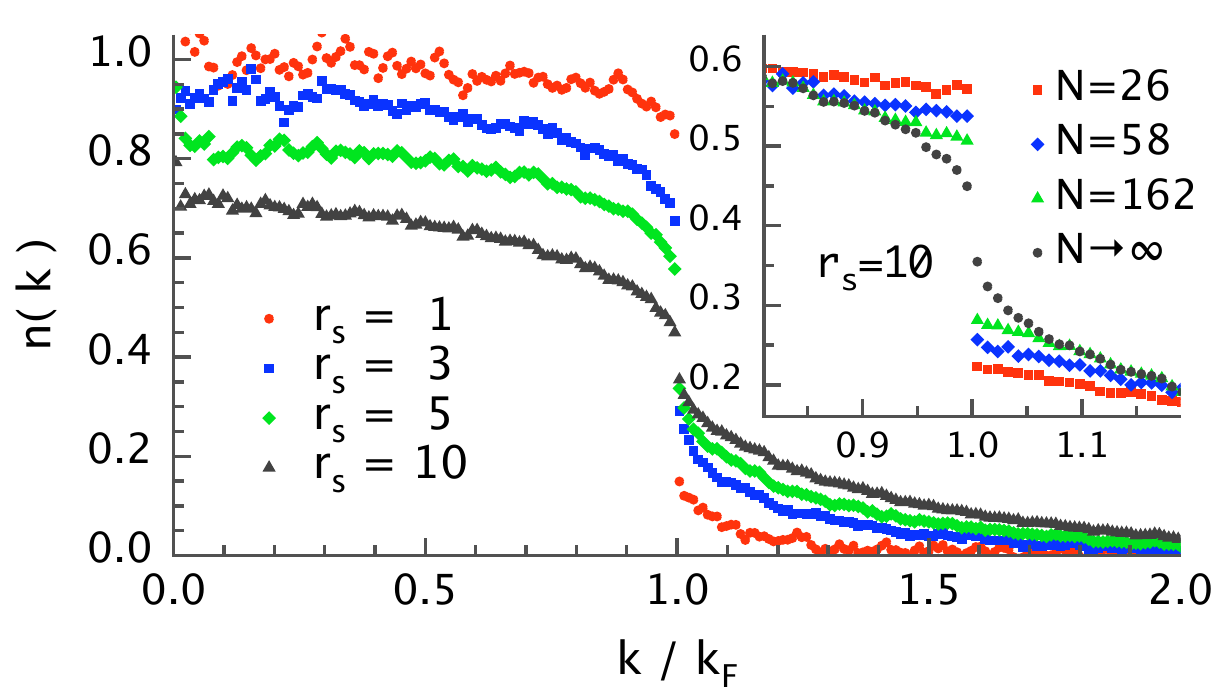}
  \caption{The momentum distribution
  of  $N=162$ unpolarized electrons using grand-canonical twist-average VMC
  for densities
  $r_s=1-10$,
  analytically corrected for size effects around $k_F$. The inset compares the uncorrected
  QMC data for different system sizes between $N=26$ and $N=162$ with the size corrected
  distribution at $r_s=10$ (``$N \to \infty$''). }
\label{MOMK}
\end{figure}

The Slater determinant of the many-body wavefunction,
Eq.~(\ref{psit}), directly connects the ground state of the
interacting system with the non-interacting one: low-lying
excitations are obtained by changing the ``occupation numbers''
of the plane waves. The energy is therefore a functional of the
occupation numbers as postulated within Landau Fermi theory
\cite{Gordon}.
Whereas this energy functional certainly exists for any finite
system, its existence in the thermodynamic limit  is
non-trivial; a necessary condition is $\lim_{N \rightarrow
\infty} Z_{k_F}^N >0$; a central issue of this paper.

We have performed Variational Quantum Monte Carlo calculations
(VMC) of the 2DEG; the electrons interact with a $1/r$
potential and with a positive background charge. We have used a
Slater-Jastrow backflow wavefunctions (SJ-BF) with an analytical form for the
both the Jastrow and backflow potentials \cite{bf};
all
potentials are split in short and long-range contributions as
described in \cite{Opt}.
For $N=58$ electrons,  the DMC ground
state energies obtained
are  $\simle 3 mRy$ lower than previous calculations using
numerically optimized
forms \cite{Kwonbf}.   Excited states were formed by adding or
subtracting orbitals in the determinant; the backflow and
Jastrow forms \cite{bf} are independent of the precise
occupation of the Slater determinant. Since the trial function
had no free parameters, we can study size effects without
re-optimizing parameters for different system sizes.

First, we calculated the momentum distribution as explained in
Ref.~\cite{Tanatar}. However, for systems in periodic boundary
conditions, the momentum distribution is only given at discrete
values $\kvec =2\pi (n \ex+m  \ey)/L$ where $n$ and $m$ are
integers and $\ex$, $\ey$ are the unit vectors in the $x$ and
$y$ direction, respectively. Using twisted boundary conditions
with twist angles $(\theta_x \ex +\theta_y \ey) 2 \pi/L$ for
the trial wavefunctions, we can obtain a momentum distribution
for all values of $\kvec$ by varying the twist angle. In the
limit of an infinite sized system, the Slater determinant of
our trial wavefunction approaches a sharp Fermi surface,
occupying only wavevectors $|\kvec| \le k_F$. For finite
systems, the sharp behavior of the occupation numbers inside
the Slater determinant is best described by working in the
grand-canonical ensemble and for a given twist angle use only
orbitals inside the Fermi sphere. This leads to a varying
particle number as a function of the twist angle. As described
in Ref.~\cite{FSE}, the translational invariance of the ground
state wavefunction allows us to define  pockets inside of which
the wavefunction transforms trivially -- any change of the
twist angle inside a pocket reduces to a  change of the total
center of mass moment accounted for by a simple phase factor;
only a single QMC calculation is needed for each pocket. As
shown in Fig.~\ref{MOMK}, the renormalization factor
quantifying the jump in the momentum distribution at $k_F$ can
be read-off precisely for any finite system. However, strong
size effects around the Fermi surface are still evident.

\begin{figure}
\includegraphics[width=0.45\textwidth]{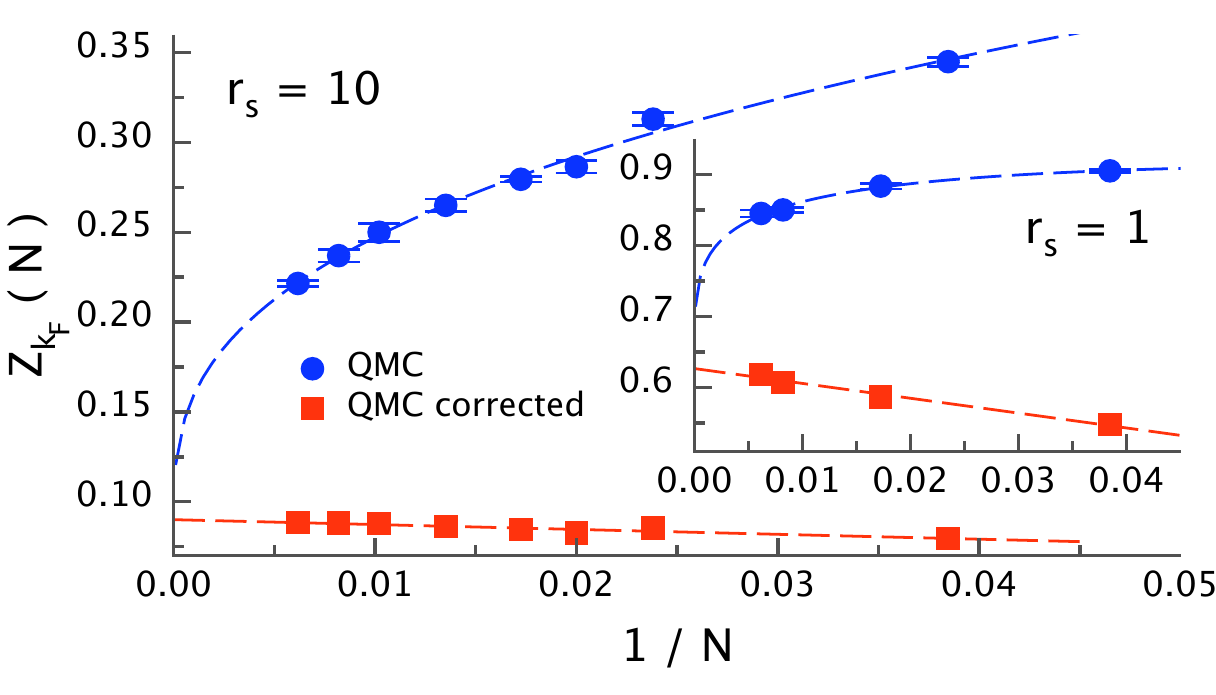}
  \caption{The renormalization factor $Z$ for $r_s=10$ estimated from the
  finite-size momentum distribution as a function of the inverse number of electrons,
  and
  the corresponding size corrected values. Dashed lines illustrate the  size corrections
  of order $N^{-1/4}$ ($N^{-1}$) for the uncorrected (corrected) data.
  The inset shows the corresponding values at
  $r_s=1$.}
\label{Z_RS}
\end{figure}

We can analyze size-effects directly using the analytical form
of the SJ-BF trial wavefunction. The
momentum distribution is obtained by displacing one particle
$\rvec_j$ a distance $\rvec$:
\begin{eqnarray}
n_\kvec^N
&=&
\left\langle e^{-i \kvec \cdot \rvec-\delta U_N}
\frac{D(\Rvec:\rvec_j+\rvec)}{D(\Rvec)} \right\rangle
\label{nkN}
\eeq
where  $\langle ... \rangle$ denotes averaging over $|\Psi_T(\Rvec)|^2$ and over a uniform
distribution for $\rvec$. The  change of the Jastrow factor  in Eq.~(\ref{nkN}) writes
\beq \delta U_N=\frac1{V}
\sum_{\qvec \ne 0}  u_q \left[ e^{i \qvec \cdot \rvec_j}
\rho_{-\qvec} -1 \right] \left[ e^{i \qvec \cdot \rvec}-1
\right] \eeq
where $\rho_\qvec=\sum_j e^{i\qvec \cdot \rvec_j}$. As described
in Ref.~\cite{FSE}, the most important finite size-effects can
be understood as an integration error by analytical
continuation of the finite-size (periodic) wavefunction to an
infinite system
where the estimator in Eq.~(\ref{nkN}) would contain the
following change in the Jastrow factor \beq
\delta U_{N\to \infty} \to \int \frac{d^2 \qvec}{(2\pi)^2} u_q \left[ e^{i \qvec \cdot \rvec_j} \rho_{-\qvec} -1 \right]
\left[ e^{i \qvec \cdot \rvec}-1 \right]. \eeq The finite-size
correction is then dominated by the nonanalyticity of the integrand
at $q = 0$
\beq
 \delta U_{\infty} -  \delta U_N & \simeq &
 \int_{-\pi/L}^{\pi/L}  \frac{d^2 \qvec}{(2\pi)^2} u_q \left[ e^{i \qvec \cdot \rvec_j} \rho_{-\qvec} -1 \right]
\left[ e^{i \qvec \cdot \rvec}-1 \right], \nonumber \eeq
and we can
calculate the leading order size corrections,
$\delta
n_\kvec \equiv  n_\kvec^\infty-n_\kvec^N$,  by expanding
$n_\kvec^\infty$, Eq.~(\ref{nkN}),
up to second order
in $\delta U_\infty-\delta U_N$.  Neglecting mode-coupling terms, we get
\beq \delta n_\kvec & \simeq &
\int_{-\pi/L}^{\pi/L}  \frac{d^2 \qvec}{(2\pi)^2} \delta(q)
\left[   n_{\kvec+\qvec}^N -n_\kvec^N \right]
\nonumber\\
&+&
\int_{-\pi/L}^{\pi/L}  \frac{d^2 \qvec}{(2\pi)^2}
 \int_{-\pi/L}^{\pi/L}  \frac{d^2 \qvec'}{(2\pi)^2}
 u_q u_{q'} \left[ 1-S(q)-S(q') \right]
 \nonumber
 \\
 &&\times
 \left[n_\kvec^N +n_{\kvec+\qvec+\qvec'}^N-n_{\kvec+\qvec}^N-n_{\kvec+\qvec'}^N
 \right]
\label{nkcorr1}
\eeq
where
\beq
\delta(q) = \left[ u_q \left(1- S(q) \right) - n u_q^2 S(q)
\right]
\label{deltaq} \eeq
Equation~(\ref{nkcorr1}) expresses
size corrections in terms of the long-wavelength limits
of the Jastrow potential and the
structure factor, $S(q)$.
For the homogenous electron gas, in the limit $q \to 0$, we have: \beq 2n u_q& \simeq &-1 +[1+
(2nv_q/\varepsilon_q)] ^{1/2} \nonumber
\\
S(q)&\simeq&[2n u_q +1/S_0(q)]^{-1}
\label{aslimits} \eeq where $v_q=2
\pi e^2/q$, $\varepsilon_q=\hbar^2 q^2/2m$, and $S_0(q)$  is the structure factor
of the non-interacting Fermi gas.

As the momentum distribution of a Fermi liquid, Eq.~(\ref{nkdef}),
is smooth everywhere away from the Fermi surface,
leading order corrections are
restricted to a small region around  $k_F$, where  we can write
 \beq \delta n_\kvec
\simeq
 Z_{k_F}^N \int_{-\pi/L}^{\pi/L} \!  \frac{d^2 \qvec}{(2\pi)^2}  \delta(q)
\left[
\theta(k_F \!-\!|\kvec+\qvec|)
\! - \!
\theta(k_F \!- \!k)
\right]. \label{deltank} \eeq
In Figure~\ref{MOMK} we show the size-corrected momentum distribution for different
densities between $r_s=1$ and $r_s=10$ using Eq.~(\ref{deltank}).
Close to $k_F$, size effects lead to important
qualitative and quantitative changes.

\begin{figure}
\includegraphics[width=0.45\textwidth]{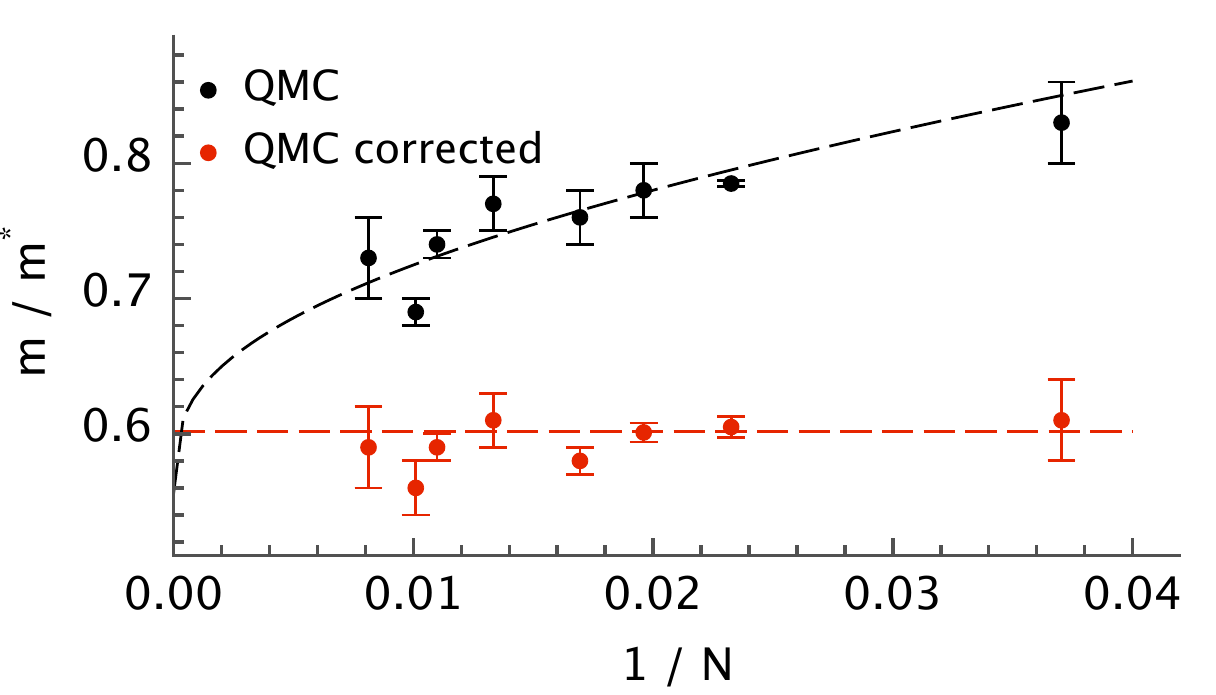}
  \caption{The inverse effective mass $m/m^*$ for $r_s=10$ as a function of $N^{-1}$, together with
  the corresponding size corrected values. Dashed lines illustrate the expected size corrections.}
\label{M_RS}
\end{figure}

The renormalization factor, $Z$, can be read-off directly from
the jump of the momentum distribution at the Fermi surface,
$Z^N_{k_F}=n_{k_F-\epsilon}^N-n_{k_F+\epsilon}^N$, and its size
corrected value may therefore be read of directly  from
Fig.~\ref{MOMK}. For a precise evaluation of $Z$ in the
thermodynamic limit, we have studied the extrapolation
separately. From  Eq.~(\ref{nkcorr1}), one can show that
size-corrections of $Z$ can be written as \beq Z_{k_F}^\infty
&\simeq& Z_{k_F}^N e^{ -\Delta_N}, \quad \Delta_N=
\int_{-\pi/L}^{\pi/L}  \frac{d^2 \qvec}{(2\pi)^2}  \delta(q)
\label{zextra} \eeq which  includes the main sub-leading order
corrections. Using the analytical forms,  Eq.~(\ref{deltaq})
and Eq.~(\ref{aslimits}), the leading order corrections are
\beq \Delta_N \simeq \left(\frac{\pi r_s^2}{4N} \right)^{-1/4}
\text{ for $N \to \infty$.} \label{LOC} \eeq The asymptotic
form, Eq.~(\ref{zextra}) with Eq.~(\ref{LOC}), shows that
actual QMC calculations with typically $N \sim 10^2$ electrons
suffer from very strong size effects. Obscured by the intrinsic
noise of QMC calculations, pure numerical analysis of the data
might suggest convergence  to values far off the exact value in
the thermodynamic limit.

In Figure \ref{Z_RS} we  compare the bare data for $r_s=1$ and
$r_s=10$ with their size corrected values. Whereas the bare
data are in reasonable  agreement with previous QMC results
\cite{Tanatar, Vignale}, a numerical extrapolation of the
uncorrected data strongly depends on assumptions on the
asymptotic scaling form, as size corrections overwhelmingly
dominate the calculation of $Z$. In order to go beyond leading
order, we have directly used Eq.~(\ref{zextra}) together with
the asymptotic forms, Eq.~(\ref{deltaq}) and
Eq.~(\ref{aslimits}), to correct our bare data analytically. As
can be seen from the figure, the size corrected values
drastically reduce size effects, as expected. More important,
in contrast to the uncorrected data, the extrapolation of the
size corrected values is not sensitive to assumptions on the
remaining corrections for densities $r_s \ge3$. 
Approaching the high density region $r_s \simle 1$, the
thermodynamic limit extrapolation is getting more difficult,
since the asymptotic expansion is singular in the limit $r_s
\to 0$. In table \ref{tableone} we have summarized our results
for the renormalization factor.

Size corrections of the momentum distribution induces size
corrections for the total kinetic energy which can be shown to
coincide with the two-dimensional analog of  Ref.~\cite{FSE}.
In two dimensions, leading order size corrections of the
kinetic and potential energy per particle scales as $N^{-5/4}$
in the Fermi liquid phase. We have added VMC and DMC energies
of the size-extrapolated values of the energy per particle in
table  \ref{tableone} .

Since our class of wavefunction  have $Z >0$, the single
particle excitation spectrum should be dominated by
quasiparticle excitations with an effective mass $m^*$. We have
calculated the effective mass by adding an electron with
momentum $\pvec$ with $|\pvec|>k_F$ to the ground state. The
effective mass of an excited state has been determined assuming
an expansion of the self energy in powers of $p-k_F$, leading
to $2m \varepsilon_p/\hbar^2= p^2-k_F^2+2 k_F (m/m^*-1)
(p-k_F)$ in the vicinity of $k_F$.

Again, the proper treatment of size effects is essential to
extrapolate to the thermodynamic limit. The additional electron
at momentum $\pvec$ will induce size corrections in the
momentum distribution which can be estimated as before. The
resulting additional finite size error in the total kinetic
energy, $\delta T_\pvec^N$, due to the excitation of momentum
$\pvec$, is then given by \beq \delta T_\pvec^N& =&
\frac{\hbar^2 \pvec^2}{2m} Z_\pvec^N \left[ e^{ -\Delta_N} -1
\right]. \label{deltaT} \eeq We see that size-corrections of
the effective mass are intrinsically related to those of the
renormalization factor, $Z$, leading to a similar asymptotic
scaling law, $N^{-1/4}$. Potential energy corrections are
independent of $\pvec$ in leading order,
 and  Eq.~(\ref{deltaT}) dominates finite size corrections for $m/m^*$. Note, that
the renormalization factor can also be obtained from analyzing
the finite-size error of effective mass calculations without
explicit calculations of the momentum distribution.

\begin{table}
\begin{tabular}{|c|c|c|c|c|c|} \hline
$r_s$  &  1 &  3  & 5 & 10 \\\hline
 $E_{\text{VMC}}  $ & -0.4179(2)  & -0.4223(1)  & -0.2975(1)  &-0.16952(1)  \\
   $E_{\text{DMC}}  $ & -0.4206(2)  &  &  -0.2991(1) & -0.17070(1) \\
$Z_{\text{VMC}}$ &  0.62(4) & 0.34(3) & 0.22(2) & 0.090(4)  \\
$ m^*/m_{\text{VMC}} $ & 1.26(7) & 1.39(8) & 1.54(7) & 1.72(9) \\\hline
$Z_{\text{RPA}}$ & 0.66 & 0.44 & 0.34 & 0.24  \\
 $m^*/m_{\text{RPA}}$ &1.02 & 1.12 & 1.16 & 1.21 \\\hline
\end{tabular}
\caption{Energies per particle (in Ry), $E_{\text{VMC}}$, and $E_{\text{DMC}}$,  the
renormalization factor $Z$ and the
effective mass $m^*/m$ extrapolated to the thermodynamic limit (both within VMC,
and from perturbative RPA calculations \cite{Schuck}).
Values in () are standard errors in the last decimal place.
} \label{tableone}
\end{table}

From figure \ref{M_RS},  we see that size effects play a
similar important role for determining $m^*$ as they do for
determining $Z$. In particular, for high densities, size
effects qualitatively change the conclusion of previous
calculations \cite{Kwon1}: whereas, in agreement with
\cite{Kwon1} all bare data indicate an effective mass smaller
than the bare mass for $N \simle 100$, in the thermodynamic
limit the effective mass is increased, as predicted by
perturbative RPA calculations \cite{Schuck,Giuliani}.

Calculations based on many-body perturbation theory going
beyond the perturbative RPA approximation have been suggested.
However,  based on different approximations, these predictions
may lead to an enhancement or depression of $Z$ (or the
effective mass) \cite{Schuck,Giuliani} and it is difficult to
estimate reliable the validity of the underlying
approximations. Our VMC results for $Z$ are always below the
corresponding values of the perturbative RPA calculations,
whereas we predict a higher effective mass $m^*/m$. Our
calculations therefore support improved RPA calculations based
on many-body local field theory  including charge- and
spin-density fluctuations as proposed in \cite{Giuliani}.

This research was supported by NSF DMR04-04853, IDRIS Computers,
and the  ACI  ``D{\`e}sordre et Interactions Coulombiennes'',
and facilitated by the Project de Collaboration CNRS/UIUC.
 M.H. thanks
S. Moroni, C. Pierleoni, R. Chitra and A. Pasturel for
discussions.

\end{document}